PACS numbers: 05.45.+b, 03.65.Ge, 03.65.Sq
------------------------------------------------------------------

\documentstyle[12pt,aps,epsf,preprint,tighten]{revtex}

\begin{document}
\draft

\title{Numerical study of scars in a chaotic billiard} 
\author{Baowen Li$^{1,2}$\footnote{E-mail: baowenli@hkbu.edu.hk; 
bwli@phchaos.hkbu.edu.hk}} 
\address{
$^{1}$ Department of Physics and Centre for Nonlinear and Complex
Systems, Hong Kong Baptist University, Hong Kong  \\
$^{2}$ Center for Applied Mathematics and Theoretical Physics,
University of Maribor, Krekova 2, 2000 Maribor, Slovenia}
\maketitle

\begin{abstract}
We study numerically the scaling properties of scars in stadium
billiard. Using the semiclassical criterion, we have searched
systematically the scars of the same type through a very wide range, from
ground state to as high as the 1 millionth state. We have analyzed the
integrated probability density along
the periodic orbit. The numerical results confirm that the average intensity 
of certain types of scars is independent of $\hbar$
rather than  scales with $\sqrt{\hbar}$. Our 
findings confirm the theoretical predictions of 
Robnik (1989).  
\end{abstract}
\pacs{PACS numbers: 05.45.+b, 03.65.Ge, 03.65.Sq}

%\maketitle
%\narrowtext
%\newpage

Eigenstates of a bound quantum system whose classical counterpart is
chaotic are of great interest in the fast developing field of quantum 
chaos. 
Among many others, scars are one of the most interesting and striking
topics. Since its discovery \cite{McDkauf,Heller84}, much
progress has been achieved.
On the theoretical side,
Bogomolny \cite{Bog88} 
developed semiclassical theory of scars in configuration space, and
Berry \cite{Berry89} performed a similar analysis in phase space
using the Wigner function. According to this theory, the intensity of 
a scar goes as $\sqrt{\hbar}$. Based on the
semiclassical evaluation of the Green function of the Schr\"odinger
equation in terms of the classical orbit, Robnik \cite{Robnik89} has 
developed a theory, and predicted that if the scar is supported by
many periodic orbits, the maximal intensity of the scar is independent of
$\hbar$, although its geometry can be determined by Bogomolny's theory. 
Most recently, Klakow and Smilansky \cite{KLSM} 
used a scattering quantization approach for this problem. Parallel to the 
theoretical developments, there have also been many numerical
\cite{WYMGR,AGFI} and experimental studies \cite{SRHE}.

Unfortunately, due to the limit of the numerical techniques and the
computer facilities, most of the numerical studies up to now are
restricted to a very low energy range, which is too low to verify 
the theoretical predictions in the very far semiclassical limit, 
particularly for Robnik's theory. In this paper, by using our numerical
code of the improved plane wave decomposition method (PWDM) (for more
details about Heller's PWDM, please see \cite{Heller90}, while for
the improved PWDM, we will discuss it in another paper\cite{Li96}), we are
successful to go as high as the 1 millionth state, which is very deep in 
the semiclassical regime for the stadium billiard.
With the help of the semiclassical criterion \cite{Robnik89},
we found many consecutive scars in several different energy ranges,
which spans 2 orders of magnitude in the wave vector, therefore, we are 
able to study the properties of scars, such as the intensity and profiles 
in the very far semiclassical limit. To the best of our knowledge, this 
is the first one of such kind of study.

To make the numerical data significant,
we need enough ensembles of scars of the same type.
Therefore, our first step is to collect scars of the same type in a 
wide range of energy. We 
begin from a very low state, e.g. the ground state. As long as we 
find the first scar, say, e.g., at wave vector $k_0$, then 
we can use the semiclassical criterion to estimate the next scar.
According to the semiclassical theory \cite{Bog88,Berry89,Robnik89}, 
the scar is most likely to occur if quantized, i.e.,

\begin{equation}
S =2\pi\hbar\left(n+\frac{\alpha}{2}\right),\qquad n=0,1,2,... .
\label{eq:action}
\end{equation}

$S$ is the action along the periodic orbit, $\alpha$ is the Maslov 
phase. It must be pointed out that
the semiclassical theory cannot predict the individual state at which 
the scar will occur. Instead, if we have already found  one scar, say, at 
$k_0$, then the semiclassical theory tells us that the eigenstates  
at the wave vector of $k_0 \pm \Delta k$
will be scarred most likely.
$\Delta k = 2\pi\hbar/{\cal L}$, ${\cal L}$ is 
the length of the periodic orbit. In our study we put $\hbar=1$, so the 
inverse of the wave vector $k$ plays the role of $\hbar$, i.e., $k$ goes to 
infinity indicates the semiclassical limit. It has been verified 
in our numerical study that this criterion is very helpful and very 
successful in searching for and collecting 
scars. As we shall see later in many cases 
this criterion is accurate within one mean level spacing, namely, the scar 
occurs at the eigenstate whose eigenenergy roughly equals the predicted 
energy. 

With the help of the semiclassical quantization criterion equation (1),
we have found about 100 examples of the same
type of scar at different energy ranges. One such example is shown in 
Fig. 1. The eigenvalue of this eigenstate is 
$k$ = 1328.153 849, which corresponds to the sequential number 250 034 for 
odd-odd parity, and to the index of about 1 001 408 when all parities are 
taken into account.  To our surprise, in addition to this one, we have 
found quite a few examples of this type of scarred state 
in such a high level. This implies that 
the scars survive the semiclassical limit. Does this finding contradict with 
the 
Shnirlman's theorem\cite{Shnirelman74}, which states that as the energy goes 
to infinity, the probability density of most eigenstates of a chaotic 
billiard approaches a uniform distribution? To test this,  we have 
investigated the statistics of the  
probability  density  distribution of the wave function, and found that it 
is an excellent Gaussian 
distribution, although there is a pronounced density around the 
periodic orbit.

In order to understand the scar properties quantitatively, we have
investigated the following pronounced (excess) intensity in a
thin tube along the periodic orbit (see Fig. 2), which is defined by
%%%%%%%%%%%%
\begin{equation}
I = \frac{\int \psi^2({\bf x}) 
d{\bf x}}{\int \left\langle\psi^2({\bf x})\right\rangle d{\bf x}} - 1.
\label{eq:int}
\end{equation}
%%%%%%%%%%%%
where $\psi({\bf x})$ is the eigenfunction at ${\bf x}$. 
$\left\langle\psi^2({\bf x})\right\rangle$ is the average probability 
density inside 
billiard, which is $1/{\cal A}$ according to the semiclassical theory 
\cite{Shnirelman74,Berry77,Voros79}. ${\cal A}$ is area of the 
billiard. 
The integral is taken over a thin tube around the periodic orbit, which 
is presented in Fig. 2.

According to Robnik's theory \cite{Robnik89}, although the geometry of a 
scar is 
determined by a single short periodic orbit, the intensity profile is 
nevertheless  determined by the sum of contributions from similar but 
longer periodic orbits, which "live" in the homoclinic neighborhood close 
to the stable and unstable manifolds of the primitive orbit. Taking into 
account all these orbits, the pronounced intensity of the scar 
defined by Eq.(\ref{eq:int}) can be described by the following formula,
%%%%%%%%%%%%
\begin{equation}
I \approx \nu \sum_{n=1}^{\infty} \frac{sin(nS_1/\hbar)}
{sinh(n\lambda\tau/2} - 1.
\label{eq:Robnik}
\end{equation}
%%%%%%%%%%%%
where, $S_1$ is the action along the primitive periodic orbit, $\lambda$ 
is the Lyapunov exponent of the primitive 
orbit with the period of $\tau$, the summation over $n$ is due to the 
repetitions of the orbit, and $\nu$ is the number of contribution orbits.  
Equation (\ref{eq:Robnik}) states that {\em the 
maximal intensity} of the scar, when supported by many periodic orbits, 
is independent of $\hbar$. This thoeretical prediction is different 
from that of Bogomolny. But, it does not contradict that of 
Bogomolny at all, instead, it is an extension of Bogomolny's theory to the 
scars  caused by many periodic orbits. These two theories describe 
different types of scars. Indeed, we have also found the scar types 
whose intensity depends on $\sqrt{\hbar}$, which is exactly predicted by 
Bogomolny's theory. However, since other authors\cite{AGFI} have 
already verified this theoretical prediction, we will not repeat this 
in this paper, we shall concentrate on the scars that cannot 
be determined by Bogomolny's 
theory, but can be described by Robnik's theory.

In Fig. 3, we show six representative examples of this intensity versus 
the width of 
the tube ($D$) in units of de Broglie wavelength around the periodic 
orbit. These six examples are the same type of 
scar, namely, the diamond-shaped scar shown in 
Fig. 1. They go from the very low state 
$k$= 10.241 095 to the very deep semiclassical regime at
$k$ = 1328.153 849. 

The first thing one can see  from these profile figures 
is that the scar intensity has a 
maximum at the width of about 1--2 de Brodglie wavelengths from the 
periodic orbit. This can be explained by Robnik's theory. The 
semiclassical waves associated with individual daughter orbits interfere 
constructively with each other only within a tube of width 1--2 de 
Brodglie wavelengths. The second important result from this figure 
is that the magnitude of the maximum does not change too much although 
the eigenenergy changes more than 100 times. 

Moreover, after checking the eigenenergies of these six examples 
carefully, we have found that 
the semiclassical criterion works very well, although we go from one scar 
state to another by jumping even up to a few hundred scarred states. 
For instance, starting from the first eigenvector $k_0$ = 10.241 095, if we 
go through 65 scarred states, we have $k = k_0 + 65\Delta k$ = 
101.563 684, 
which is very close to the exact one $k_{exact}$ =101.568 640. ( 
In this paper, we study only the eigenstates with odd-odd 
parity, so the length of the periodic orbit  shown in Fig. 2 is
${\cal L} = 2\sqrt{5}$, rather than $4\sqrt{5}$ for the total billiard,
thus, $\Delta k = 
2\pi/{\cal L}$ = 1.404 96). The deviation is less than one mean level 
spacing.
This procedure applies also to many other scarred states and it can be 
verified readily for other states given in Fig. 3.

The next very important question is that how does this  
maximal integrated intensity depend on the eigenenergy 
or the $\hbar$?  Firstly, our numerical results show that 
around a certain $k$, it 
changes from scarred state to state. This is shown in Fig. 4, where we 
plot 26 consecutive 
scarred states around $k=125$. (Note that there are two cases in which two 
consecutive eigenstates are near degenerate, thus both of them are 
scarred.)  Again,  
from this figure we can see clearly that the  semiclassical 
criterion (\ref{eq:action}) works very excellently. The interval between two 
scarred 
states is almost constant and approximately equals $2\pi/{\cal L}$. The 
maximal integrated intensity, however, fluctuates from state to state, 
which cannot be explained by the existing semiclassical approaches. 
This is still an open problem deserving of further theoretical  
and numerical investigations.

The results shown in Fig. 4 imply that
in order to make the study of the dependence of the maximal
integrated intensity 
on energy significant we should take certain kinds of ensemble averaging.
In our numerical study, we 
have performed such averaging around 
certain $k$ over 
many scarred states (usually about 10 states).  The averaged data are 
plotted in Fig. 5. The best least-squares fit gives rise to
%%%%%%%%%%%%
\begin{equation}
\left\langle I_{m}\right\rangle = 0.73/k^a,\qquad a = 0.06\pm 0.03,
\end{equation}
%%%%%%%%%%%%
here, $\left\langle \right\rangle$ means the local average. The exponent 
$a$ is very close to 
zero and is far from $1/2$ as predicted by Bogomolny's theory. This fact 
means that the maximal integrated intensity does not depend on the energy or 
$\hbar$ for the scar type shown in this paper. This discovery is very 
different from previous ones \cite{AGFI} 
and cannot be explained by the semiclassical theory of Bogomolny
\cite{Bog88} and Berry \cite{Berry89}, however, it confirms 
quantitatively the theoretical prediction of 
Robnik \cite{Robnik89}, which states that the maximal intensity of a 
scar is 
independent of $\hbar$ if the scar is supported by many orbits, as 
mentioned above.

In this paper, we have studied intensively the scars in a stadium billiard, 
and have shown numerically that the semiclassical criterion 
(\ref{eq:action}) works very well from very low state to that in the very 
far 
semiclassical limit. Furthermore, we have analyzed the scaling property 
of a scar with $\hbar$ and found that for the scar type shown in this 
paper, the maximal integrated density fluctuates from scarred state to
state, but the local average intensity does not change with energy.
This finding confirms Robnik's scar theory of multiple periodic orbits 
\cite{Robnik89}. 
\bigskip 
\\

The author would like to thank Professor Dr. Marko Robnik for 
discussions. He is also very  grateful to Professor Dr. Felix Izrailev for 
helpful  discussions during the STATPHYS 19 in Xiamen (1995) and during 
his visit in Como (1996). This work was supported in part by the Research 
Grant Council Grant
RGC/96-97/10 and the Hong Kong Baptist University Faculty Research Grants
FRG/95-96/II-09 and FRG/95-96/II-92. The work done in Slovenia was 
supported by the Ministry of Science and Technology of Republic of Slovenia.

 \newpage
%%
%
%\end{document}
%
\newpage
\clearpage

\begin{figure}
%\vspace{-1.cm}
%\centerline{\epsfxsize 15cm \epsffile{scarfig1.ps}}
\caption{The probability density plot for a scarred eigenstate of 
odd-odd parity. The wave number $k$ = 1328.153 849, which corresponds to 
index 250 034 using the Weyl formula (odd-odd), thus it corresponds to 
approximately the 1 001 408th eigenstate for the total billiard. The 
scar is obviously supported by the diamond-shaped periodic orbit shown in 
Fig. 2. The stadium has the parameter of circle
radius R = 1 and the straight line length 2. In this figure, the unit length
is about 211 de Broglie wavelengths.}
\end{figure}

\begin{figure}
%\vspace{-30mm}
%\centerline{\epsfxsize 14cm \epsffile{scarfig2.ps}}
%\vspace{-20mm}
\caption{The integral region around the periodic orbit that is taken in 
Eq. (2). The width of the tube is 
$D$  measured perpendicular to the periodic orbit.} 
\end{figure}

\begin{figure}
%\centerline {\epsfxsize 16cm \epsffile{scarfig3.ps}}
%\vspace{-15mm}
\caption{The integrated scar intensity $I$ 
vs the width of 
the integrating tube in unit of the de Broglie wavelength for the scar 
type shown in Fig. 1. We show six different eigenstates at different 
energy ranges.} 
\end{figure}

\begin{figure}
%\centerline {\epsfxsize 14cm \epsffile{scarfig4.ps}}
%\vspace{-15mm}
\caption{
The maximum of integrated scar 
intensity vs wave number $k$ around $k=125$. The type of scar 
is the same as shown in Fig. 1 (the diamond shape). Here we
see clearly that the wave number interval between two consecutive scars 
is very close to $2\pi/{\cal L}$ (= 1.404 96), as predicted by the 
semiclassical quantization condition Eq.(\ref{eq:action}).}
\end{figure}

\begin{figure}
%\centerline {\epsfxsize 14cm \epsffile{scarfig5.ps}}
%\vspace{-15mm}
\caption{The locally averaged (over a small group of 
consecutive scarred states) maximum of integrated scar excess intensity
vs wave number $k$. The bullet represents the numerical data, and the 
solid line is the best least-squares fit.}
\end{figure}
\end{document}